\documentclass[aps,prb,twocolumn,longbibliography]{revtex4-2}

\usepackage{physics}
\usepackage{tikz} 
\usetikzlibrary{shapes,arrows,positioning,automata,backgrounds,calc,er,patterns}
\usepackage{graphicx}
\usepackage{dcolumn}
\usepackage{bm}
\usepackage{amsmath}
\usepackage{amssymb}
\usepackage{amsthm}
\usepackage{amsfonts}
\usepackage{enumerate}
\usepackage{siunitx}
\usepackage{slashed}
\usepackage{makecell}

\usepackage{epstopdf}

\usepackage[colorlinks=true,urlcolor=blue,citecolor=blue,linkcolor=blue]{hyperref}

\usepackage{CJK} 

\definecolor{purple}{rgb}{0.8,0,0.6}
\definecolor{darkgreen}{rgb}{0.00,0.6,0.00}

\newcommand{\affiliationSICNU}{Department of Physics, Institute of Solid State Physics and Center for Computational Sciences, Sichuan Normal University, Chengdu, Sichuan 610066, China}

\begin{document}
	\begin{CJK*}{UTF8}{}
		\title{Weyl-mediated Ruderman-Kittel-Kasuya-Yosida interaction revisited: imaginary-time formalism and finite temperature effects}
		
		\author{Mengyao Zhou}\affiliation{\affiliationSICNU}
		
		\author{Hao-Ran Chang ({\CJKfamily{gbsn}{张浩然}})
		}\affiliation{\affiliationSICNU}
		\author{Lijun Yang}\email{yanglijun@sicnu.edu.cn}
		\affiliation{\affiliationSICNU}
		\author{Long Liang}\email{longliang@sicnu.edu.cn}
		\affiliation{\affiliationSICNU}

		\begin{abstract}
			Noncentrosymmetric magnetic Weyl semimetals provide a platform for investigating the interplay among magnetism, inversion symmetry breaking, and topologically non-trivial Weyl fermions. 
			The Weyl-mediated Ruderman-Kittel-Kasuya-Yosida (RKKY) interaction may be related to the magnetic orders observed in rare-earth magnetic Weyl semimetals.
			Previous studies of RKKY interaction between magnetic impurities in Weyl semimetals found Heisenberg, Ising-like, and  Dzyaloshinskii-Moriya~(DM) types of interactions. However, different range functions are obtained in the literature. 
			In this work, we calculate the Weyl-mediated RKKY interaction by using the divergence-free imaginary-time formalism and obtain exact analytical results at finite temperature. The discrepancies among zero-temperature range functions in the literature are resolved.
			At nonzero temperature, the interaction strength decays exponentially in the long distance limit.
			But in the short distance limit, the DM interaction shows a thermal enhancement, an effect persists up to higher temperature for shorter distance. This provides a mechanism stabilizing the helical order observed in rare-earth magnetic Weyl semimetals.
		\end{abstract}
		\maketitle
		
	\end{CJK*}
	
	\section{Introduction}
	Weyl semimetals represent an exotic state of matter and have attracted a lot of research interest~\cite{Armitage:rev-2018,HongDingRMP}. 
	The emergence of Weyl fermions requires the breaking of either inversion~\cite{ShuichiMurakami_2007} symmetry or time-reversal~\cite{PhysRevB.83.205101} symmetry. The latter is usually related to the presence of magnetic orders. Weyl semimetals that break both time-reversal and inversion symmetries have been predicted in various materials, such as Heusler magnets~\cite{YanSun1,YanSun2,Heusler} and rare-earth compounds  $R$Al$X$ where $R$=rare-earth and $X$=Ge or Si~\cite{PhysRevB.97.041104}. Recently, materials in the rare-earth family have been experimentally identified~\cite{Sanchez2020}, providing a platform for exploring various fascinating phenomena 
	arising from the interplay among magnetism, inversion symmetry breaking, and topologically nontrivial electronic states~\cite{Sanchez2020,PhysRevLett.124.017202,Yang_2020,Gaudet2021,PhysRevB.103.115143,PhysRevX.13.011035,Drucker2023,PhysRevMaterials.7.034202,PhysRevB.108.205143,PhysRevB.109.035120,PhysRevB.109.195130,PhysRevX.14.021012,PhysRevB.109.035143,PhysRevB.110.224436,2503.14768}.
	
	The $R$Al$X$ compounds show complex magnetic orders, including ferromagnetism~\cite{Sanchez2020,Yang_2020}, antimerons~\cite{PhysRevLett.124.017202}, helimagnetic  order~\cite{Gaudet2021}, non-collinear ferromagnetism~\cite{PhysRevB.103.115143}, antiferromagnetism~\cite{PhysRevB.109.035120}  and multi-$Q$ order~\cite{2503.14768}. 
	Despite experimental complexity, a common feature in different materials is the emergence of helical order~\cite{Gaudet2021,PhysRevX.13.011035,Drucker2023}, with the momentum coincident with the nesting of Weyl nodes, indicating possible connections between magnetism and Weyl fermions. 
	
	The mechanism behind the magnetic orders and phase transitions between the orders is still under investigation. It was suggested that 
	the Weyl-mediated Ruderman-Kittel-Kasuya-Yosida~(RKKY) interactions~\cite{RK,K,Y} provide a predominant mechanism for the emergence of helical order~\cite{Gaudet2021,PhysRevX.13.011035,2412.20743}. The Kondo coupling between Weyl fermions and local magnetic moments can also stabilize the helical order~\cite{Drucker2023,2502.03832}. 
	A recent first-principle study proposed a different scenario, in which the magnetic order originates from frustrated superexchange interaction between local magnetic moments~\cite{PhysRevB.109.L201108}.  
	
	The Weyl-mediated RKKY interactions have been extensively studied in the past decade~\cite{PhysRevB.92.241103,PhysRevB.92.224435,PhysRevB.93.094433,Sun_2017,PhysRevB.96.115204,Duan_2018,PhysRevB.99.165111,PhysRevB.99.165302,WANG2020126327,PhysRevB.103.155151,Kundu_2023,PhysRevB.107.165147}. However, although the same types of interactions, i.e., the Heisenberg, link-dependent Ising, and Dzyaloshinskii-Moriya~(DM)~\cite{DZYALOSHINSKY1958,Moriya1960} terms were discovered,  different forms of the range functions were obtained in the literature. Most previous studies reported $1/r^5$ or $1/r^3$ decay of range functions in the long distance limit, depending on whether the chemical potential is vanishing or not. But a $1/r^2$ decay was obtained in~\cite{PhysRevB.103.155151}, which also predicted non-universal oscillation periods depending on a high energy cutoff. 
	
	Given that the Weyl-meditated RKKY interaction is a fundamental problem and may play a pivotal role in understanding the magnetic properties of Weyl semimetals, it becomes an important task to clarify the behavior of range functions and resolve discrepancies in the previous literature. Furthermore, the observation of thermally robust helical domains above the thermodynamic magnetic transition temperature~\cite{Drucker2023} demands an investigation of finite temperature effects on the Weyl-mediated RKKY interaction. 
	
	In this work, we reinvestigate the problem of Weyl-mediated RKKY interactions. The real-time calculations for RKKY interaction  suffer from the ultraviolet divergence originating from the linear dispersion relation, just as in the case of graphene~\cite{PhysRevB.76.184430}. To clarify the origin of divergence, we compare the imaginary- and real-time formalisms and arrive at the condition
	for their equivalence. Within the divergence-free imaginary-time formalism~\cite{PhysRevB.58.3584,PhysRevB.84.115119}, we obtain exact analytical expressions of the range functions at finite temperature, which automatically reduce to their zero temperature counterparts reported in~\cite{PhysRevB.92.241103}. We further reveal the reason behind the discrepancies between our results and the findings in~\cite{PhysRevB.103.155151}. In addition, at finite temperature, the range functions decay exponentially in the long distance limit. More importantly, we find that in the short distance limit, the DM interaction in the same chirality channel is remarkably enhanced by temperature. The thermal enhancement of DM interaction provides a mechanism that stabilizes the thermally robust helical order~\cite{Drucker2023}.
	
	The rest of this paper is organized as follows. In Section~\ref{Sec:method}, we derive the RKKY interaction using the imaginary-time formalism and discuss its connection with real-time formalism. In Section~\ref{Section:results}, we calculate the Weyl-mediated RKKY interaction at finite temperature and exact analytical expressions are obtained. In Section~\ref{Section:zeroT}, we analyze the zero temperature results and point out the reason behind the discrepancies in the literature. Section~\ref{Section:finiteT} focuses on the finite temperature effects with an emphasis on the thermal enhancement of DM interaction.  Section~\ref{Section:conclusion} is devoted to the conclusion. Some detailed derivations, comparisons, and lattice model calculations that confirm our analytical results are presented in the appendixes.
	
	\section{Method}\label{Sec:method}
	In this section we first give a brief derivation of the RKKY interaction in the imaginary-time formalism~\cite{PhysRevB.58.3584,PhysRevB.84.115119}, and then discuss its connection with real-time expressions used in the literature. The two formalisms are equivalent only under certain assumption.
	
	We consider two magnetic impurities $\mathbf{S}_1$ and $\mathbf{S}_2$ located at $\mathbf{r}_1$ and $\mathbf{r}_2$ interact with the itinerate electrons through the Kondo coupling. The Hamiltonian for the system is 
	\begin{eqnarray}
		H=H_0+J \mathbf{S}_1\cdot\psi^\dag(\mathbf{r}_1)\bm{\sigma}\psi(\mathbf{r}_1)+J\mathbf{S}_2\cdot\psi^\dag(\mathbf{r}_2)\bm{\sigma}\psi(\mathbf{r}_2),
	\end{eqnarray}
	where $H_0$ is the Hamiltonian for itinerate electrons, $J$ is the Kondo coupling constant,  $\psi^\dag(\mathbf{r})=\{\psi^\dag_{\uparrow}(\mathbf{r}),\psi^\dag_{\downarrow}(\mathbf{r})\}$ with $\psi^\dag_{\uparrow/\downarrow}(\mathbf{r})$ being the creation operator for spin-up or spin-down electron at position $\mathbf{r}$, and $\bm{\sigma}=\{\sigma_{x},\sigma_{y},\sigma_{z}\}$ are the Pauli matrices. We take the reduced Planck constant $\hbar$ and Boltzmann constant $k_B$ to be unity throughout this work.
	The partition function is 
	\begin{eqnarray}
		\mathcal{Z}=\int \mathcal{D}(\psi,\bar{\psi})e^{-\int^{1/T}_0\mathrm{d}\tau (\bar{\psi}\partial_\tau \psi+H)}=e^{-(\Omega_0+\Delta\Omega)/T},
	\end{eqnarray}
	where $T$ is the temperature, $\Omega_0$ is the grand potential in the absence of magnetic impurities, and $\Delta \Omega$ is the correction to the grand potential due to magnetic impurities. 
	Up to the lowest order of $J$, $\Delta \Omega$ is given by~\cite{PhysRevB.84.115119}
	\begin{widetext}
		\begin{eqnarray}
			\Delta \Omega=J^2\sum_{\alpha\beta}\int \mathrm{d}\tau~\mathrm{tr}[\sigma_\alpha G(\tau,\mathbf{r}_1-\mathbf{r}_2)\sigma_\beta G(-\tau,\mathbf{r}_2-\mathbf{r}_1)]S_{1\alpha}(\mathbf{r}_1)S_{2\beta}(\mathbf{r}_2),
			\label{Eq:DeltaOmega}
		\end{eqnarray}
	\end{widetext}
	where $\mathrm{tr}$ represents the trace over spin degrees of freedom and
	$G(\tau_1,\mathbf{r}_1-\mathbf{r}_2)$ is the non-perturbed imaginary time Green's function of itinerant electrons. 
	The change in the grand potential, Eq.~\eqref{Eq:DeltaOmega}, can be understood as an effective interaction between the local moments induced by itinerate electrons, known as the RKKY interaction,  $H_{\mathrm{RKKY}}=\Delta\Omega=\sum_{\alpha\beta}\mathcal{J}_{\alpha\beta}(\mathbf{r}_1-\mathbf{r}_2)S_{1\alpha}S_{2\beta}$. In the Matsubara frequency representation~\cite{PhysRevB.58.3584,PhysRevB.84.115119},  
	\begin{eqnarray}\label{Eq:RKKY_Matsubara}
		\mathcal{J}_{\alpha\beta}(\mathbf{r})=TJ^2\sum_{\omega_n}
		\mathrm{tr} [\sigma_\alpha G(i\omega_n,\mathbf{r})\sigma_\beta G(i\omega_n,-\mathbf{r})],
	\end{eqnarray}
	with  $\omega_n=(2n+1)\pi T$ being the fermionic Matsubara frequency.

	In the following we show that, under certain assumption, the real-time expressions can be derived from the imaginary-time result. To this end, we first write the Green's function by using the spectral representation~\cite{altland_simons_2010},
	\begin{eqnarray}
		G(i\omega_n,\mathbf{r})=\int\frac{\mathrm{d}z}{2\pi}\frac{A(z,\mathbf{r})}{i\omega_n-z},
	\end{eqnarray}
	where $A(z,\mathbf{r})$ is the spectral function of itinerant electrons. 
	Using the spectral representation, Eq.~\eqref{Eq:RKKY_Matsubara} can be rewritten as
	\begin{eqnarray}\label{Eq:RKKY_Matsubara2}
		\mathcal{J}_{\alpha\beta}(\mathbf{r})=TJ^2\sum_{\omega_n}\int\frac{\mathrm{d}z}{2\pi}\int\frac{\mathrm{d}z'}{2\pi}
		\frac{\rho_{\alpha\beta}(z,z',\mathbf{r})}{(i\omega_n-z)(i\omega_n-z')},
	\end{eqnarray}
	with $\rho_{\alpha\beta}(z,z',\mathbf{r})=\mathrm{tr}[\sigma_\alpha A(z,\mathbf{r})\sigma_\beta A(z',-\mathbf{r})]$. 
	
	Note that in Eq.~\eqref{Eq:RKKY_Matsubara2}, integrations over $z$ and $z'$ should be calculated before the Matsubara frequency summation over $i\omega_{n}$. However, suppose that  the result is unchanged if we change the order of the Matsubara frequency summation over $i\omega_{n}$ and the integrations over $z$ and $z^\prime$ [this is true, for example, if Eq.~\eqref{Eq:RKKY_Matsubara2} converges absolutely], after performing the frequency summation  with the help of the residues theorem, we obtain,
	\begin{widetext}
		\begin{eqnarray}
			\mathcal{J}_{\alpha\beta}(\mathbf{r})=-J^2\int\frac{\mathrm{d}z}{2\pi}\int\frac{\mathrm{d}z'}{2\pi}
			\int \frac{\mathrm{d}\omega}{2\pi i}
			\bigg[\frac{1}{(\omega-z+i0^+)(\omega-z'+i0^+)}-\frac{1}{(\omega-z-i0^+)(\omega-z'-i0^+)}\bigg]f(\omega)\rho_{\alpha\beta}(z,z^\prime,\mathbf{r}),\label{Eq:RKKY_real1}
		\end{eqnarray}
		where $f(\omega)=1/(e^{\omega/T}+1)$ is the Fermi-Dirac distribution.
		Reordering the integration order and integrating over $z$ and $z^\prime$ first, Eq.~\eqref{Eq:RKKY_real1} becomes
		\begin{eqnarray}
			\mathcal{J}_{\alpha\beta}(\mathbf{r})=-J^2\int \frac{\mathrm{d}\omega}{2\pi i}
			\mathrm{tr} [\sigma_\alpha G^r(\omega,\mathbf{r})\sigma_\beta G^r(\omega,-\mathbf{r})-\sigma_\alpha G^a(\omega,\mathbf{r})\sigma_\beta G^a(\omega,-\mathbf{r})]f(\omega),\label{Eq:RKKY_real2}
		\end{eqnarray}
	\end{widetext}
	where $G^r(\omega,\mathbf{r})$ and $G^a(\omega,\mathbf{r})$ are the retarded and advanced Green's functions, respectively. 
	Using the relation $G^a(\omega,\mathbf{r})=G^{r\dag}(\omega,-\mathbf{r})$ and the cyclic property of the trace,  we have $\mathrm{tr}[\sigma_\alpha G^a(\omega,\mathbf{r})\sigma_\beta G^a(\omega,-\mathbf{r})]=\mathrm{tr}[\sigma_\alpha G^r(\omega,\mathbf{r})\sigma_\beta G^r(\omega,-\mathbf{r})]^\ast$. As a consequence, 
	Eq.~\eqref{Eq:RKKY_real2} is simplified as
	\begin{eqnarray}
		\mathcal{J}_{\alpha\beta}(\mathbf{r})=-J^2\mathrm{Im}\int \frac{\mathrm{d}\omega}{\pi}
		\mathrm{tr}[\sigma_\alpha G^r(\omega,\mathbf{r})\sigma_\beta G^r(\omega,-\mathbf{r})]f(\omega),\label{Eq:RKKY_real}
	\end{eqnarray}
	which is the real-time form widely used in the literature.  
	
	From the above analyses, it becomes clear that, the real-time formalism is equivalent to the imaginary-time one, provided that it is legitimate to switch the order of Matsubara frequency summation and integration of spectral function. For lattice systems, this is always allowed as the result is finite. However, in the continuum descriptions of Weyl semimetals and graphene, the low energy theories are governed by linear dispersion relations, and the
	frequency integration in Eq.~\eqref{Eq:RKKY_real} diverges. To obtain finite results, proper regularization is required~\cite{PhysRevB.76.184430,PhysRevB.92.241103,PhysRevB.92.224435}.
	In contrast, the imaginary-time formalism gives finite results~\cite{PhysRevB.84.115119}. Thus in this work, we utilize  Eq.~\eqref{Eq:RKKY_Matsubara} to calculate the Weyl-mediated RKKY interaction.

	In~\cite{PhysRevB.103.155151}, the RKKY interaction is calculated by performing the Fourier transform of the spin-spin correlation function  in the  momentum space.
	To understand the connection between that approach and our result, we integrate out $\omega$ in Eq.~\eqref{Eq:RKKY_real1} and the result is,
	\begin{eqnarray}
		\mathcal{J}_{\alpha\beta}(\mathbf{r})=J^2\int\frac{\mathrm{d}z}{2\pi}\int\frac{\mathrm{d}z'}{2\pi}\frac{f(z')-f(z)}{z'-z}\rho_{\alpha\beta}(z,z^\prime,\mathbf{r}).\label{Eq:RKKY_real3}
	\end{eqnarray}
	Expressing $\rho_{\alpha\beta}(z,z^\prime,\mathbf{r})$ in Eq.~\eqref{Eq:RKKY_real3} as the Fourier transform of $\rho_{\alpha\beta}(z,z^\prime,\mathbf{q})$, we arrive at the expression for the spin-spin correlation function in the momentum space~\cite{PhysRevB.103.155151}. 
	However, the momentum space  correlation function diverges, and an ultraviolet cutoff was introduced to obtain finite results in~\cite{PhysRevB.103.155151}.

	\section{model and results}\label{Section:results}
	The low energy Hamiltonian for a Weyl semimetal with $N$ nodes is
	\begin{eqnarray}
		H=\sum^N_{a=1}\sum_{\mathbf{k}} \psi^\dag_{\mathbf{Q}_a+\mathbf{k}} (\chi_a v_F \mathbf{k}\cdot\bm{\sigma}+\epsilon_{a} )\psi_{\mathbf{Q}_a+\mathbf{k}},\label{Eq:HWeyl}
	\end{eqnarray}
	here $\chi_a=\pm 1$ denotes the chirality of the $a$-th Weyl node,  $v_F$ is the Fermi velocity, $\mathbf{Q}_a$ is the location of the $a$-th node in momentum space, $\mathbf{k}$ is measured from $\mathbf{Q}_a$, and $\epsilon_{a}$ is the energy of the $a$-th Weyl node. In this work, we assume that the Weyl dispersion is spherically symmetric.
	Note that in Eq.~\eqref{Eq:HWeyl}, $\bm{\sigma}$ denotes pseudo-spin, 	which is a mixture of spin and orbit degrees of freedom. The relation between real spin and pseudospin depends on material details. As a minimal model, we here assume that the pseudo-spin is the same as the real spin, an approximation widely used in the literature. 
	
	The Weyl-mediated RKKY interaction can be written as $\mathcal{J}_{\alpha\beta}(\mathbf{r})=\sum_{ab}\mathcal{J}^{ab}_{\alpha\beta}(\mathbf{r})$, where
	\begin{eqnarray}
		\mathcal{J}^{ab}_{\alpha\beta}(\mathbf{r})=TJ^2\sum_{\omega_n}
		\mathrm{tr} [\sigma_\alpha G_a(i\omega_n,\mathbf{r})\sigma_\beta G_b(i\omega_n,-\mathbf{r})],\label{Eq:Jab}
	\end{eqnarray}
	describes the interaction mediated by nodes $a$ and $b$. 
	
	To proceed, we first calculate the real space Green's function for the $a$-th node,
	\begin{eqnarray}
		G_a(i\omega_n,\mathbf{r})=\frac{e^{i\mathbf{Q}_a\cdot\mathbf{r}}}{2}\sum_{s=\pm}\int\frac{\mathrm{d}^3\mathbf{k}}{(2\pi)^3}\frac{(1+s\chi_a\hat{\mathbf{k}}\cdot\bm{\sigma})e^{i\mathbf{k}\cdot\mathbf{r}}}{i\omega_n-s v_F k+\mu_a},~
	\end{eqnarray}
	here $\mu_a=\mu-\epsilon_{a}$ is the effective chemical potential for the $a$-node, $\hat{\mathbf{k}}=\mathbf{k}/k$ and $s=\pm$ denotes particle and hole bands. 
	To calculate the Fourier transform, we write $\hat{\mathbf{k}}$ as $-i\bm{\nabla}_\mathbf{r}/k$, and then it is straightforward to compute the integral in spherical coordinates. The result is (see Appendix~\ref{Append:details} for details)
	\begin{eqnarray}
		G_a(i\omega_n,\mathbf{r})
		&=&-\frac{(i\omega_n+\mu_a)(1+\chi_ag_a\bm{\sigma}\cdot\hat{\mathbf{r}})}{4\pi v^3_F\tilde{r}}e^{i\phi},\label{Eq:Ga}
	\end{eqnarray}
	where $\hat{\mathbf{r}}=\mathbf{r}/r$, $\tilde{r}=r/v_F$, $g_a=\mathrm{sgn}(\omega_n)+i/[(i\omega_n+\mu_a)\tilde{r}]$, and $\phi=\mathbf{Q}_a\cdot\mathbf{r}+\mathrm{sgn}(\omega_n)(i\omega_n+\mu_a)\tilde{r}$ with $\mathrm{sgn}$ being the sign function. 
	It follows  that the Matsubara Green's function is suppressed exponentially with increasing $|\omega_n|$, thus the frequency summation in Eq.~\eqref{Eq:Jab} converges. 
	
	Substituting Eq.~\eqref{Eq:Ga} into Eq.~\eqref{Eq:Jab}, after some algebra, we find that the RKKY interaction takes the following form,
	\begin{eqnarray}
		\mathcal{J}^{ab}_{\alpha\beta}(\mathbf{r})
		=\mathcal{J}^{ab}_{\mathrm{H}}(\mathbf{r})\delta_{\alpha\beta}+\mathcal{J}^{ab}_{\mathrm{I}}(\mathbf{r})\hat{r}_{\alpha}\hat{r}_\beta+\mathcal{J}^{ab}_{\mathrm{DM}}(\mathbf{r})\epsilon_{\alpha\beta\gamma}\hat{r}_\gamma.~
	\end{eqnarray}	
	Physically, $\mathcal{J}^{ab}_{\mathrm{H}}(\mathbf{r})$ gives the isotropic Heisenberg interaction, $\mathcal{J}^{ab}_{\mathrm{I}}(\mathbf{r})$ describes a link-dependent Ising-like interaction, and $\mathcal{J}^{ab}_{\mathrm{DM}}(\mathbf{r})$ leads to a DM interaction 
	$\mathbf{D}_{ij}\cdot\mathbf{S}_i\times\mathbf{S}_j$
	with the vector $\mathbf{D}_{ij}$ being parallel to $\mathbf{r}_i-\mathbf{r}_j$. Explicitly, the exact analytical expressions of range functions at finite temperature are (see Appendix.~\ref{Append:details} for details).
	
	\begin{widetext}
		\begin{eqnarray}
			&&\mathcal{J}^{ab}_{\mathrm{H}}
			=
			\frac{C}{\tilde{r}^2}\bigg[\cos{(2\bar{\mu}_{ab}\tilde{r})}
			\frac{(1+\chi_a\chi_b)\tilde{r}^2(\mu_b\mu_b-\partial^2_{2\tilde{r}})+\chi_a\chi_b(\tilde{r}\partial_{\tilde{r}}-1)}{\tilde{r}^2}
			+2
			\bar{\mu}_{ab}\sin{(2\bar{\mu}_{ab}\tilde{r})}		
			\frac{\tilde{r}(1+\chi_a\chi_b)\partial_{2\tilde{r}}-\chi_a\chi_b}{\tilde{r}}
			\bigg]F,~~~\label{Eq:JH}\\
			&&\mathcal{J}^{ab}_{\mathrm{I}}
			=
			\frac{2\chi_a\chi_bC}{\tilde{r}^2}\bigg[\cos{(2\bar{\mu}_{ab}\tilde{r})}
			\frac{\tilde{r}^2(\partial^2_{2\tilde{r}}-\mu_b\mu_b)+1-\tilde{r}\partial_{\tilde{r}}}{\tilde{r}^2}
			+2
			\bar{\mu}_{ab}\sin{(2\bar{\mu}_{ab}\tilde{r})}
			\frac{1-\tilde{r}\partial_{2\tilde{r}}}{\tilde{r}}
			\bigg]F,\label{Eq:JI}\\
			&&	\mathcal{J}^{ab}_{\mathrm{DM}}
			=
			\frac{C}{\tilde{r}^2}\bigg[\cos{(2\bar{\mu}_{ab}\tilde{r})}
			\frac{\chi_b\mu_a+\chi_a\mu_b-\tilde{r}(\chi_a+\chi_b)\bar{\mu}_{ab}\partial_{\tilde{r}}}{\tilde{r}}
			+(\chi_a+\chi_b)
			\sin{(2\bar{\mu}_{ab}\tilde{r})}
			\frac{\tilde{r}\mu_a\mu_b+\partial_{2\tilde{r}}-\tilde{r}\partial^2_{2\tilde{r}}}{\tilde{r}}
			\bigg]F,\label{Eq:JDM}
		\end{eqnarray}
		where  $C=J^2\cos{(\mathbf{Q}_{ab}\cdot\mathbf{r})}/(8\pi^2v^6_F)$
		with $\mathbf{Q}_{ab}=\mathbf{Q}_{a}-\mathbf{Q}_{b}$ being the separation of Weyl nodes in the momentum space, 
		$\bar{\mu}_{ab}=(\mu_a+\mu_b)/2$ is the average effective chemical potential, and $\partial_{\tilde{r}}=\partial/\partial \tilde{r}$.  The temperature dependence is encoded in  
		$F=T\csch{(2\pi T\tilde{r})}$. 
		The range functions oscillate with a period determined by the chemical potential as well as energies of the Weyl nodes. The oscillation is additionally modulated by $\mathbf{Q}_{ab}$. Because of this modulation, the RKKY interaction may prefer magnetic orders with momentum around the Weyl nodes separation.

		\subsection{Zero temperature limit}\label{Section:zeroT}
		In the zero temperature limit, the range functions reduce to
		\begin{eqnarray}
			&&\mathcal{J}^{ab}_{\mathrm{H}}
			=
			\frac{C}{2\pi}
			\bigg[\frac{\cos{(2\bar{\mu}_{ab}\tilde{r})}}{\tilde{r}^3}
			\bigg((1+\chi_a\chi_b)\mu_b\mu_b-\frac{1+5\chi_a\chi_b}{2\tilde{r}^2}
			\bigg)
			-
			\frac{\sin{(2\bar{\mu}_{ab}\tilde{r})}}{\tilde{r}^4}(1+3\chi_a\chi_b)\bar{\mu}_{ab}\bigg],\label{Eq:JH_T0}\\
			&&\mathcal{J}^{ab}_{\mathrm{I}}
			=
			\frac{\chi_a\chi_b C}{\pi}
			\bigg[\frac{\cos{(2\bar{\mu}_{ab}\tilde{r})}}{\tilde{r}^3}
			\bigg(
			\frac{5}{2\tilde{r}^2}-\mu_b\mu_b
			\bigg)
			+\frac{3\sin{(2\bar{\mu}_{ab}\tilde{r})}\bar{\mu}_{ab}}{ \tilde{r}^4}
			\bigg],\label{Eq:JI_T0}\\
			&&\mathcal{J}^{ab}_{\mathrm{DM}}
			=
			\frac{C}{2\pi}
			\bigg[
			\frac{\cos{(2\bar{\mu}_{ab}\tilde{r})}}{\tilde{r}^4}
			[(\chi_a+\chi_b)\bar{\mu}_{ab}+\chi_b\mu_a+\chi_a\mu_b
			]+
			\frac{(\chi_a+\chi_b)\sin{(2\bar{\mu}_{ab}\tilde{r})}}{\tilde{r}^3}
			\bigg(\mu_a\mu_b-\frac{1}{\tilde{r}^2}
			\bigg)
			\bigg],\label{Eq:JDM_T0}
		\end{eqnarray}
		which are consistent with the results obtained in~\cite{PhysRevB.92.241103}. 
		Note that $\mu_a\mu_b$ is proportional to the geometric mean of the densities of states of Weyl fermions around nodes $a$ and $b$, thus the amplitude of the RKKY interactions becomes larger with increasing $\mu_a$ and $\mu_b$.
		However, even for $\mu_a=\mu_b=0$, the interaction is still nonvanishing, and the contribution comes from interband transitions instead of the Fermi surface.

		\begin{table*}[h]
			\caption{Decay law of the range functions at zero temperature in this work.}
			\label{Table:decay_zero_temperature}
			\begin{tabular}{c|c|c}
				\hline\hline
				& Long distance                                                                                                                     & Short distance                                                                                                   \\ \hline
				Heisenberg & \makecell{$1/r^3$ (doped, same chirality)\\ $1/r^4$ (doped, opposite chirality)\\ $1/r^5$ (undoped)} & $1/r^5$                                                                                                          \\ \hline
				Ising     & \makecell{$1/r^3$ (doped)\\ $1/r^5$ (undoped)}                                                       & $1/r^5$                                                                                                          \\ \hline
				DM        & \makecell{$1/r^3$ (doped, same chirality)\\ $1/r^4$ (opposite chirality, $\mu_a\ne \mu_b$)}           & \makecell{$1/r^5$ (same chirality)\\ $1/r^4$ (opposite chirality, $\mu_a\ne \mu_b$)} \\ \hline\hline
			\end{tabular}
		\end{table*}
	\end{widetext}

	In Table~\ref{Table:decay_zero_temperature}, we show the decay laws of range functions Eqs.~\eqref{Eq:JH_T0}-\eqref{Eq:JDM_T0} in both the long- and short-distance limits. 
		Our zero temperature results 
		are consistent with most previous studies summarized in Table~\ref{Table:literature} (see Appendix~\ref{Append:Tables}) but differ from those of~\cite{PhysRevB.103.155151} in two critical ways. Firstly, in the large $r$ limit, the range functions, Eqs.~\eqref{Eq:JH_T0}-\eqref{Eq:JDM_T0}, decay as $1/r^n$ with  $n=3$, 4, or 5, depending on the chiralities of the nodes, the chemical potential, and the energies of the Weyl nodes. By contrast, all range functions decay as $1/r^2$ for large $r$ in~\cite{PhysRevB.103.155151}. Secondly, apart from the modulation by $\mathbf{Q}_{ab}$,  the period of the range functions in this work is determined by $\bar{\mu}_{ab}$. In the case of $\mu_a=\mu_b=\mu$, $\bar{\mu}_{ab}\tilde{r}=\mu \tilde{r}=k_F r$, thus the period is the same as that of the conventional RKKY interaction. But in~\cite{PhysRevB.103.155151}, the oscillation period depends on a high energy cutoff, which is to some extent arbitrary. In particular, for $\mu_a=\mu_b$, apart from the $\cos{(\mathbf{Q}_{ab}\cdot\mathbf{r})}$ modulation, our results predict a $1/r^5$ decay without oscillation of Heisenberg and Ising-like couplings; however, the range functions in~\cite{PhysRevB.103.155151} always oscillate.

	\begin{figure*}
		\includegraphics[width=0.9\textwidth]{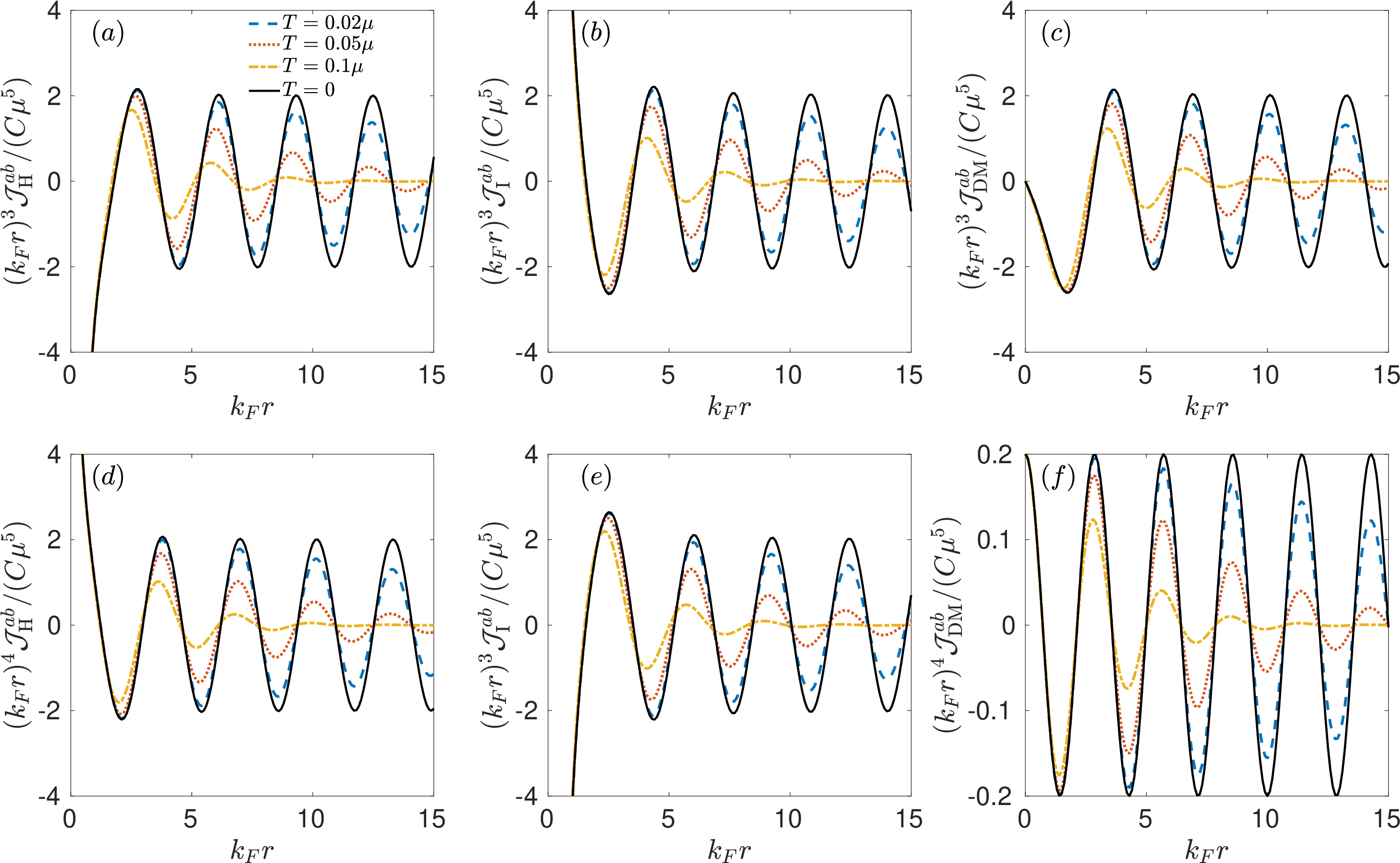}
		\caption{Range functions as functions of $k_F r=\mu\tilde{r}$ at different temperatures. In (a)-(c), $\chi_a=\chi_b=1$, and in (d)-(f), $\chi_a=-\chi_b=1$. We take $\mu_a=\mu_b=\mu$ in (a)-(e), while in (f) $\mu_a=\mu$ and  $\mu_b=1.2\mu$ are used since the DM interaction mediated by opposite chirality Weyl fermions vanishes when $\mu_a=\mu_b$.
		}\label{Fig:figure1}
	\end{figure*}

	The discrepancies originate from the fact that, the RKKY interaction in~\cite{PhysRevB.103.155151} was calculated from the Fourier transform of the momentum space spin-spin correlation function. But the latter diverges, and an ultraviolet momentum cutoff $\Lambda$ was introduced to get finite results. 
	To expose the flaws in this approach, we perform the Fourier transform of the range functions, Eqs.~\eqref{Eq:JH_T0}-\eqref{Eq:JDM_T0}. The results are shown in Appendix~\ref{Append:Fourier transform}.
	We find that the Fourier transform diverges, and if a short distance cutoff $r\sim 1/\Lambda$ is introduced, the leading divergence is proportional to  $\Lambda^2$, which is the same as the result in~\cite{PhysRevB.103.155151}. Now if we keep only the leading divergent term and try to recover the range functions by an inverse Fourier transform, we will get the specious $1/r^2$ decay and cutoff-dependent oscillations as obtained in~\cite{PhysRevB.103.155151}. 
	Our analyses thus show that, it is not a proper way to calculate the range function  through the Fourier transform from the momentum space result if the latter diverges. This is not unexpected, since mathematically, the Fourier transform of a function does not necessarily exist.

	To further confirm the validity of our results, we calculate the RKKY interaction by using a lattice model with two Weyl nodes, see Appendix~\ref{Append:lattice model}. The numerical results are in good agreement with our analytical predictions.

	\subsection{Finite temperature effects}\label{Section:finiteT}
	
	Now we investigate the effects of nonzero temperature. In the long distance limit ($T\tilde{r}>>1$), $F\approx T e^{-2\pi T\tilde{r}}$, and therefore, in contrast to the algebraic decay at zero temperature, the range functions decay exponentially for large $r$ at nonvanishing temperature~\cite{PhysRevB.51.670}. 
	
	Figure~\ref{Fig:figure1} shows the dimensionless range functions as functions of $k_F r=\mu\tilde{r}$ for the same [(a)-(c)] and opposite chirality channels [(d)-(f)] at several different temperatures. The chemical potential is taken to be the energy unit. 
	In panels (a)-(e) we take $\mu_a=\mu_b=\mu$. To get a nonvanishing DM interaction in the opposite chirality channel [cf. Eq.~\eqref{Eq:JDM}], $\mu_a=\mu$ and $\mu_b=1.2\mu$ in (f) are used.
	For clarity, the results are scaled by $(k_Fr)^3$ [(a)-(c) and (e)] or $(k_Fr)^4$ [(d) and (f)] so that in the large $r$ limit at zero temperature the amplitudes of the oscillations approach to constants. As evidenced by the plots, the amplitudes of the interactions decrease with increasing temperature and distance. For the Heisenberg and DM interactions with $\mu_a\mu_b\ne 0$ and in the long distance limit, the leading contribution in the same chirality channel originates from the Fermi surface and decays as $1/r^3$; while in the opposite chirality channel, the range functions decays as $1/r^4$.
	So the  Heisenberg and DM interactions of the same chirality channel are generally larger than those of the opposite chirality channel  [note that different scaling factors are used in panels (a)/(c) compared to (d)/(f)]. 
	In contrast, for the Ising-like interaction, the same- and opposite-chirality channels yield results of equal magnitude but opposite sign, cf. Eq.~\eqref{Eq:JI}.
	
	\begin{figure*}
		\includegraphics[width=0.9\textwidth]{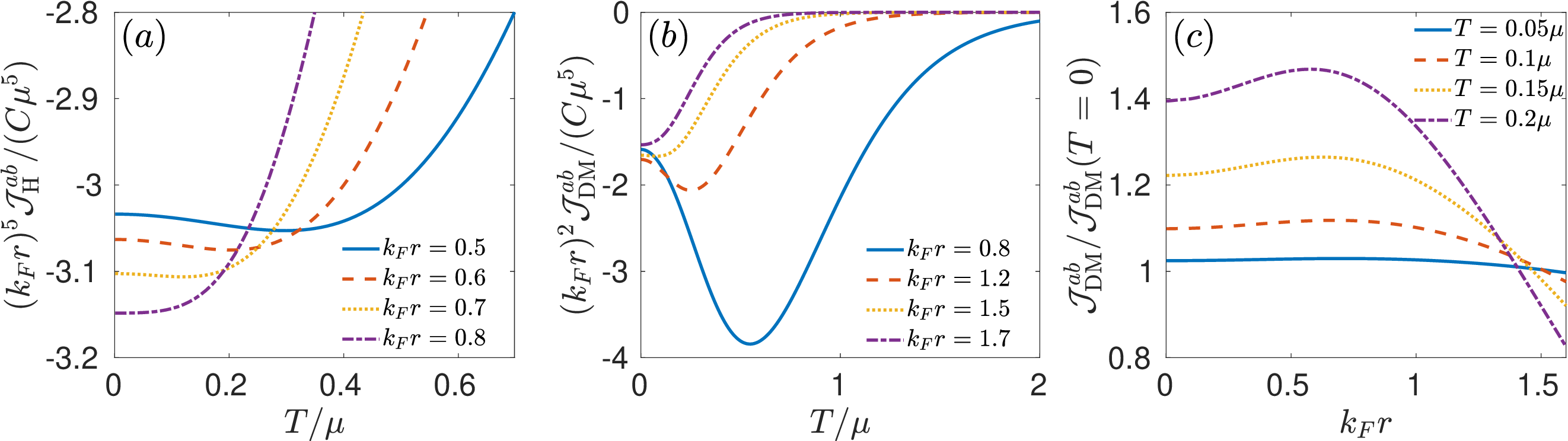}
		\caption{The dimensionless Heisenberg (a) and DM (b) range functions in the same chirality ($\chi_a=\chi_b=1$) channel as functions of temperature for different $k_F r$. Panel (c) shows the ratio of finite and zero temperature DM interactions as a function of $k_F r$. We take $\mu_a=\mu_b=\mu$ as energy unit. 
		}\label{Fig:figure2}
	\end{figure*}
	
	Since the interactions become short-ranged at finite temperature, in the following we focus on the short distance behavior of the range functions.
	We find that, surprisingly, the Heisenberg and DM interactions in the same chirality channel can increase with increasing temperature in certain parameter regime, while other range functions decrease slightly with increasing temperature in the short distance and low temperature limit. Keeping the leading order in $r$ and $T$, the same chirality ($\chi_a=\chi_b=\chi$) Heisenberg and DM range functions in the small $r$ and $T$ limit are 
	\begin{eqnarray}
		&&J^{ab}_{\mathrm{H}}=
		-\frac{C}{6\pi}
		\frac{9+4\mu_a\mu_b T^2\pi^2\tilde{r}^4}{\tilde{r}^5},\label{Eq:J_H_low_r_T} \\
		&&J^{ab}_{\mathrm{DM}}=
		-\frac{\chi C}{3\pi}
		\frac{\mu^3_a+\mu^3_b+2\bar{\mu}_{ab}\pi^2T^2}{\tilde{r}^2}.\label{Eq:J_DM_low_r_T}
	\end{eqnarray}
	The above results indicate that, with increasing temperature, the DM interaction increases since $\bar{\mu}_{ab}$ and $\mu^3_a+\mu^3_b$ take the same sign, and the Heisenberg interaction increases if $\mu_a\mu_b>0$.

	Figure~\ref{Fig:figure2}~(a) and (b)  show the dimensionless Heisenberg and DM interaction strengths as functions of temperature at several different $k_F r$. The Heisenberg interaction is scaled by $(k_F r)^5$ and the DM interaction is scaled by $(k_F r)^2$. Figure~\ref{Fig:figure2}~(c) shows 
	the DM interaction strength normalized by the zero temperature value as a function of $k_F r$ at several temperatures. 
	As shown in the figure, the DM interaction exhibits a more pronounced temperature-induced enhancement compared to the Heisenberg counterpart. 
	For $k_F r$ below a critical value, the interaction strength first increases with temperature, consistent with Eqs.~\eqref{Eq:J_H_low_r_T} and \eqref{Eq:J_DM_low_r_T}. As we mentioned before, the range functions decrease exponentially with temperature in the high temperature limit. Thus the interaction strength shows a non-monotonic feature with increasing temperature for small $k_F r$, as demonstrated  in Fig.~\ref{Fig:figure2}~(a) and (b) . 
	The thermal enhancement persists up to higher temperature for smaller $k_F r$.  If $k_F r$ exceeds a critical value, the interaction strength shows a monotonic decrease with rising temperature. The Fermi momentum is typically of the order of $0.1\pi/a$~\cite{2412.20743} (where $a$ is the lattice constant), thus the DM interaction between local moments separated by up to several lattice constants could be enhanced by increasing temperature.  
	A thermal enhancement of the RKKY interaction strength was also founded in graphene at charge neutral point~\cite{PhysRevB.92.205414}. This is different from our finding, since at charge neutrality, the Weyl-mediated DM interaction vanishes identically.
	
	Our starting point is a low energy effective model, Eq.~\eqref{Eq:HWeyl}, and in principle the results are valid in the long range limit. To verify the validity of the short distance expansion, we calculate the DM interaction using a lattice model in Appendix~\ref{Append:lattice model}, and a more profound thermal enhancement is observed.
	
	The	Weyl semimetals $R$Al$X$ exhibit complex magnetic orders as varying temperature and external magnetic field. However, a common feature is the presence of a helical order whose momentum coincides with the nesting vector between Weyl nodes. The helical order is thermally robust, and nanoscale magnetic domains were observed above the thermodynamic transition temperature~\cite{Drucker2023}. The physical mechanism behind this phenomenon is under current investigation~\cite{PhysRevB.109.L201108,2412.20743,2502.03832}. Nevertheless, regardless of its origin, the helical order could be stabilized by thermal enhancement of the Weyl-mediated DM interaction.

	The competition between thermal fluctuations and the thermally enhanced DM
		interaction qualitatively explains the stabilization of magnetic domains within a temperature window,
		as observed experimentally~\cite{Drucker2023}. 
		Since the thermal enhancement occurs in the same chirality channel, the momentum of the helical order is expected to be close the the separation between Weyl nodes with the same chirality.
		Furthermore, note that the momentum of the helical
		order is related to the DM interaction strength~\cite{P_Bak_1980}. Therefore our result indicates that, the momentum of the helical order increases with raising temperature. This prediction can be verified using neutron scattering experiments.
	
	\section{Conclusion and discussion}\label{Section:conclusion}
	
	In this work, we reinvestigate the RKKY interaction mediated by Weyl fermions. Utilizing the imaginary-time formalism, we obtain analytical results at finite temperature. 
	We resolve the discrepancies among zero temperature results in the literature.  At nonvanishing temperature,  range functions are exponentially suppressed with increasing distance. More importantly, we find that in the short distance limit, the DM interaction in the same chirality channel increases with temperature in certain parameter regime. This thermal enhancement could stabilize the helical magnetic order observed in rare-earth magnetic Weyl semimetals.  
	
	To better elucidate the Weyl-mediated RKKY interactions in real materials, the effects of Weyl cone tilt and the difference between pseudo-spin and real spin should be taken into account. Due to the spin-orbit coupling, the pseudospin consists of both spin and orbital degrees of freedom. Thus the Kondo coupling, when projected to the pseudopsin space, can become anisotropic and node-dependent~\cite{PhysRevB.92.241103}. While our results can be generalized to capture the node-dependent Kondo coupling, the anisotropic Kondo coupling may introduce additional directional dependencies that require further exploration.
	Previous studies~\cite{PhysRevB.99.165111,2412.20743} show that the tilt can induce extra DM interactions with the DM vector perpendicular to the direction of impurity displacement. However, the precise form of these extra interactions and their temperature dependence demands further investigation.  These are left for future work.

	\acknowledgments
	We are grateful to Xi Luo for discussions and comments. 
	This work was supported by the National Natural Science
	Foundation of China under Grants No.~12204329, No.~12204331, No.~11547200, and the Research Institute of Intelligent Manufacturing Industry Technology of  Sichuan Arts and Science University.

	\appendix
	\section{Detailed derivations of the RKKY interaction}\label{Append:details}
	In this appendix, we provide detailed derivations of the RKKY interactions. We first calculate the real space Matsubara Green's function for the $a$-th node ($\mu_a=\mu-\epsilon_{a}$),
		\begin{eqnarray}
			G_a(i\omega_n,\mathbf{r})=\frac{e^{i\mathbf{Q}_a\cdot\mathbf{r}}}{2}\sum_{s=\pm}\int\frac{\mathrm{d}^3\mathbf{k}}{(2\pi)^3}\frac{(1+s\chi_a\hat{\mathbf{k}}\cdot\bm{\sigma})e^{i\mathbf{k}\cdot\mathbf{r}}}{i\omega_n-s v_F k+\mu_a},~~
		\end{eqnarray}
		to calculate the integral we write $\hat{\mathbf{k}}=-i\bm{\nabla}_{\mathbf{r}}/k$,  After integrating out the  azimuthal and polar angles in spherical coordinates and summing over the band indices, we get
		\begin{widetext}
			\begin{eqnarray} 
				G_a(i\omega_n,\mathbf{r})
				=\frac{e^{i\mathbf{Q}_a\cdot\mathbf{r}}}{2}\int^\infty_{-\infty}\frac{k\mathrm{d}k}{2\pi^2}\frac{1-i\chi_a/k\bm{\sigma}\cdot\hat{\mathbf{r}}\partial_r}{i\omega_n-v_F k+\mu_a}\frac{\sin{(kr)}}{r}.
			\end{eqnarray}
			The integral can be calculated by using the residue theorem. Introducing $\tilde{r}=r/v_F$, the result is,
			\begin{eqnarray} 
				G_a(i\omega_n,\mathbf{r})&=&-\frac{e^{i\mathbf{Q}_a\cdot\mathbf{r}}[(\mu_a+i\omega_n) -i\chi_a\bm{\sigma}\cdot\hat{\mathbf{r}}\partial_{\tilde{r}} ]}{4\pi v^3_F}\bigg[\frac{e^{i(\mu_a\tilde{r}+i\omega_n)\tilde{r}}\theta(\omega_n)+e^{-i(\mu_{a}\tilde{r}+i\omega_n)\tilde{r}}\theta(-\omega_n)}{\tilde{r}}\bigg],\\
				&=&-\frac{e^{i\mathbf{Q}_a\cdot\mathbf{r}}(\mu_a+i\omega_n)}{4\pi v^3_F}\bigg[1+\chi_a\bm{\sigma}\cdot\hat{\mathbf{r}}\bigg(\mathrm{sgn}(\omega_n)+\frac{i}{(\mu_a+i\omega_n)\tilde{r}}\bigg)\bigg ]\frac{e^{i~\mathrm{sgn}(\omega_n)(\mu_a+i\omega_n)\tilde{r}}}{\tilde{r}},\\
				&\equiv&-\frac{e^{i\mathbf{Q}_a\cdot\mathbf{r}}(\mu_a+i\omega_n)}{4\pi v^3_F}\bigg[1+\chi_ag_a\bm{\sigma}\cdot\hat{\mathbf{r}}\bigg ]\frac{e^{i~\mathrm{sgn}(\omega_n)(\mu_a+i\omega_n)\tilde{r}}}{\tilde{r}}.
			\end{eqnarray}
			The RKKY interaction mediated by Weyl fermions around nodes $a$ and $b$ is [$\mathbf{Q}_{ab}=\mathbf{Q}_a-\mathbf{Q}_b$ and $\bar{\mu}_{ab}=(\mu_a+\mu_b)/2$],
			\begin{eqnarray}\label{Eq:append1_J}
				\mathcal{J}^{ab}_{\alpha\beta}(\mathbf{r})
				&=&\frac{TJ^2\cos{(\mathbf{Q}_{ab}\cdot\mathbf{r}})}{8\pi^2v^6_F \tilde{r}^2}\sum_{\omega_n}(i\omega_n+\mu_a)(i\omega_n+\mu_b)
				e^{2i~\mathrm{sgn}(\omega_n)(\bar{\mu}_{ab}+i\omega_n)\tilde{r}}\nonumber\\
				&&
				[\delta_{\alpha\beta}(1+\chi_a\chi_bg_ag_b)-2\chi_a\chi_bg_ag_b\hat{r}_\alpha\hat{r}_\beta
				-i\epsilon_{\alpha\beta\gamma}(\chi_ag_a+\chi_bg_b)\hat{r}_\gamma],
			\end{eqnarray}
			Note that the full interaction consists of sum over $a$ and $b$, so only terms that are symmetric in $a$ and $b$ contribute.
			To get the above result we have used the identity
			\begin{eqnarray}
				\mathrm{tr}[\sigma_\alpha (1+\mathbf{A}\cdot \bm{\sigma})\sigma_\beta (1+\mathbf{B}\cdot\bm{\sigma})] = 2 [\delta_{\alpha\beta}(1-\mathbf{A}\cdot\mathbf{B})+A_\alpha B_\beta+A_\beta B_\alpha-i\epsilon_{\alpha\beta\gamma}(A_\gamma-B_\gamma)].
			\end{eqnarray}
			
			Form Eq.~\eqref{Eq:append1_J}, we can see that the interaction takes the form
			\begin{eqnarray}
				\mathcal{J}^{ab}_{\alpha\beta}(\mathbf{r})
				&=&\mathcal{J}^{ab}_{\mathrm{H}}(\mathbf{r})\delta_{\alpha\beta}+\mathcal{J}^{ab}_{\mathrm{I}}(\mathbf{r})\hat{r}_{\alpha}\hat{r}_\beta+\mathcal{J}^{ab}_{\mathrm{DM}}(\mathbf{r})\epsilon_{\alpha\beta\gamma}\hat{r}_\gamma,
			\end{eqnarray}
			where\begin{eqnarray}
				\mathcal{J}^{ab}_{\mathrm{H}}(\mathbf{r})
				=\frac{TJ^2\cos{(\mathbf{Q}_{ab}\cdot\mathbf{r})}}{8\pi^2v^6_F \tilde{r}^2}\sum_{\omega_n}(i\omega_n+\mu_{a})(i\omega_n+\mu_{b})
				(1+\chi_a\chi_bg_ag_b)e^{2i~\mathrm{sgn}(\omega_n)(\bar{\mu}_{ab}+i\omega_n)\tilde{r}},
			\end{eqnarray}
			gives an isotropic Heisenberg interaction,
			\begin{eqnarray}
				\mathcal{J}^{ab}_{\mathrm{I}}(\mathbf{r})
				&=&-\frac{TJ^2\cos{(\mathbf{Q}_{ab}\cdot\mathbf{r})}}{4\pi^2v^6_F \tilde{r}^2}\sum_{\omega_n}(i\omega_n+\mu_{a})(i\omega_n+\mu_{b})
				\chi_a\chi_bg_ag_be^{2i~\mathrm{sgn}(\omega_n)(\bar{\mu}_{ab}+i\omega_n)\tilde{r}},
			\end{eqnarray}
			describes an Ising-like coupling, and
			\begin{eqnarray}
				\mathcal{J}^{ab}_{\mathrm{DM}}(\mathbf{r})
				&=&-\frac{iTJ^2\cos{(\mathbf{Q}_{ab}\cdot\mathbf{r})}}{8\pi^2v^6_F \tilde{r}^2}\sum_{\omega_n}(i\omega_n+\mu_{a})(i\omega_n+\mu_{b})
				(\chi_ag_a+\chi_bg_b)e^{2i~\mathrm{sgn}(\omega_n)(\bar{\mu}_{ab}+i\omega_n)\tilde{r}},
			\end{eqnarray}
			is a DM interaction. Using
			\begin{eqnarray}
				T\sum_{\omega_n>0}e^{-2\omega_n \tilde{r}}&=&T\sum_{\omega_n<0}e^{2\omega_n\tilde{r}}=\frac{T\csch{(2\pi T\tilde{r}})}{2},
			\end{eqnarray}
			the Mastubara frequency summation can be easily calculated, and we obtain the results in the paper.


			\section{Decay laws and parameters controlling the oscillation period of the range functions in Weyl-mediated RKKY interaction}\label{Append:Tables}
			
			To confirm our zero-temperature results and to show the main discrepancies between our results and Ref.~\cite{PhysRevB.103.155151}, we summarize the decay laws and  parameters controlling the oscillation period of the range function reported previously ~\cite{PhysRevB.92.241103,PhysRevB.92.224435,PhysRevB.93.094433,Sun_2017,PhysRevB.96.115204,Duan_2018,PhysRevB.99.165111,PhysRevB.99.165302,WANG2020126327,PhysRevB.103.155151} in Table~\ref{Table:literature}. It is emphasized that the results in the literature ~\cite{PhysRevB.92.241103,PhysRevB.92.224435,PhysRevB.93.094433,Sun_2017,PhysRevB.96.115204,Duan_2018,PhysRevB.99.165111,PhysRevB.99.165302,WANG2020126327,PhysRevB.103.155151} were restricted to the zero-temperature limit.
			
			\begin{table*}[h]
				\caption{Decay laws and parameters controlling oscillation period of the range functions reported previously.}
				\label{Table:literature}
				\begin{tabular}{c|c|c|c}
					\hline\hline
					& System   & \makecell{Decay laws  in the \\long-distance limit }                                                                                                          & \makecell{Parameters controlling \\the oscillation period }                                                                                   \\ \hline
					Refs.~\cite{PhysRevB.92.241103,PhysRevB.92.224435} & Weyl/Dirac semimetal  &\makecell{$1/r^3$ (doped)\\  $1/r^5$ (undoped)} & \makecell{nodes separation\\  Fermi momentum} 
					\\ \hline                                        
					Ref.~\cite{PhysRevB.93.094433} & Dirac semimetal &\makecell{$1/r^3$ (doped)\\ } & \makecell{nodes separation\\  chemical potential\\
						Fermi velocity anisotropy}                                                \\ \hline
					Ref.~\cite{Duan_2018} &\makecell{Weyl semimetal \\one impurity is on the surface\\the other one is in the bulk} &\makecell{$1/r^3$ (doped, bulk)\\ $1/r^5$ (undoped, bulk)\\ 
						$1/r^2$ (doped, surface effect)\\ $1/r^3$ (undoped, surface effect)} & \makecell{ nodes separation\\  chemical potential\\
						Fermi velocity anisotropy}                                              \\ \hline
					Refs.~\cite{PhysRevB.99.165111, WANG2020126327} & tilted Weyl/Dirac &\makecell{$1/r^3$ (doped)\\ 
						$1/r^5$ (undoped)} &\makecell{
						nodes separation\\  Fermi momentum\\tilt}                                  \\ \hline
					Ref.~\cite{PhysRevB.99.165302}  &\makecell{Dirac semimetal\\two impurities on the surface
					} &\makecell{$1/r^3$ (doped, bulk effect)\\ $1/r^5$ (undoped, bulk effect)\\ 
						$1/r$ or $1/r^4$ (surface effect)} & \makecell{nodes separation\\  Fermi momentum}                                           \\ \hline
					Ref.~\cite{PhysRevB.103.155151}    & Weyl semimetal     & $1/r^2$          & \makecell{cutoff\\
						nodes separation\\  Fermi momentum }   \\ \hline\hline
				\end{tabular}
			\end{table*}

		\section{Fourier transform of the range functions}\label{Append:Fourier transform}
		Here we calculate the Fourier transform of the range functions. 
		In the following, the momentum $\mathbf{p}$ is measured from $\mathbf{Q}_{ab}$ such that the $\cos{(\mathbf{Q}_{ab}\cdot\mathbf{r})}$ modulation in the range functions can be neglected. In the momentum space,
		\begin{eqnarray}
			\mathcal{J}^{ab}_{\alpha\beta}(\mathbf{p})&=&\int\mathrm{d}^3\mathbf{r}\bigg[\mathcal{J}^{ab}_{\mathrm{H}}(r)\delta_{\alpha\beta}+\mathcal{J}^{ab}_{\mathrm{I}}(r)\hat{R}_{\alpha}\hat{r}_\beta+\mathcal{J}^{ab}_{\mathrm{DM}}(r)\epsilon_{\alpha\beta\gamma}\hat{r}_\gamma\bigg]e^{-i\mathbf{p}\cdot\mathbf{r}},\\
			&=&4\pi\int  r^2\mathrm{d}r\bigg[\mathcal{J}^{ab}_{\mathrm{H}}(r)\delta_{\alpha\beta}-\mathcal{J}^{ab}_{\mathrm{I}}(r)/r^2\hat{p}_\alpha \hat{p}_\beta\partial^2_{p}+i\mathcal{J}^{ab}_{\mathrm{DM}}(r)/r\epsilon_{\alpha\beta\gamma}\hat{p}_\gamma\partial_{p}\bigg]\frac{\sin{(p r)}}{pr},\\
			&\equiv&\mathcal{J}^{ab}_{\mathrm{H}}(p)\delta_{\alpha\beta}+\mathcal{J}^{ab}_{\mathrm{I}}(p)\hat{p}_\alpha \hat{p}_\beta+i\epsilon_{\alpha\beta\gamma}\mathcal{J}^{ab}_{\mathrm{DM}}(p)\hat{p}_\gamma,
		\end{eqnarray}
		where $\partial_p=\partial/\partial p$ and
		\begin{eqnarray}
			&&\mathcal{J}^{ab}_{H}(p)=4\pi\int \mathrm{d}r~r\mathcal{J}^{ab}_{\mathrm{H}}(r)\frac{\sin{(p r)}}{p},\\
			&&\mathcal{J}^{ab}_{\mathrm{I}}(p)
			=-4\pi\partial^2_p\int \mathrm{d}r~\frac{\mathcal{J}^{ab}_{\mathrm{I}}(r)}{r}\frac{\sin{(p r)}}{p},\\
			&&\mathcal{J}^{ab}_{\mathrm{DM}}(p)
			=4\pi\partial_p\int \mathrm{d}r~\mathcal{J}^{ab}_{\mathrm{DM}}(r)\frac{\sin{(p r)}}{p}.
		\end{eqnarray}
		The above integrals diverge in the $r\to 0$ limit, to get finite results we introduce a short-distance cutoff $\tilde{r}=1/\Lambda$, and then in the zero-temperature limit, 
		\begin{eqnarray}
			\mathcal{J}^{ab}_{\mathrm{H}}(p)
			&=&\frac{J^2}{4\pi^2v^3_F p}\bigg[
			\bigg(\frac{1+5\chi_a\chi_b}{2}I^{cs}_4-
			(1+\chi_a\chi_b)\mu_b\mu_bI^{cs}_2
			\bigg)
			+(1+3\chi_a\chi_b)\bar{\mu}_{ab}I^{ss}_3
			\bigg]\\,
			\mathcal{J}^{ab}_{{\mathrm{I}}}(p)
			&=&\frac{J^2\chi_a\chi_b}{2\pi^2v^3_F }\partial^2_p\frac{1}{p}\bigg[
			\bigg(\frac{5}{2}I^{cs}_6-
			\mu_b\mu_bI^{cs}_4
			\bigg)
			+3\bar{\mu}_{ab}I^{ss}_5
			\bigg],\\
			\mathcal{J}^{ab}_{{\mathrm{DM}}}(p)
			&=&\frac{J^2}{4\pi^2v^3_F}\partial_p\frac{1}{p}\bigg[
			(\chi_a+\chi_b)(I^{ss}_5-\mu_a\mu_b I^{ss}_3)-[(\chi_a+\chi_b)\bar{\mu}_{ab}+\chi_b\mu_a+\chi_a\mu_b
			]I^{cs}_4
			\bigg],
		\end{eqnarray}
		where 
		\begin{eqnarray}
			&&I^{cs}_n=\int^\infty_{1/\Lambda} \mathrm{d}\tilde{r}
			\frac{\cos{(2\bar{\mu}_{ab}\tilde{r})}\sin{(v_Fp\tilde{r}})}{\tilde{r}^n},~~
			I^{cc}_n=\int^\infty_{1/\Lambda} \mathrm{d}\tilde{r}
			\frac{\cos{(2\bar{\mu}_{ab}\tilde{r})}\cos{(v_Fp\tilde{r}})}{\tilde{r}^n},\\
			&&I^{ss}_n=\int^\infty_{1/\Lambda} \mathrm{d}\tilde{r}
			\frac{\sin{(2\bar{\mu}_{ab}\tilde{r})}\sin{(v_Fp\tilde{r}})}{\tilde{r}^n},~~
			I^{sc}_n=\int^\infty_{1/\Lambda} \mathrm{d}\tilde{r}
			\frac{\sin{(2\bar{\mu}_{ab}\tilde{r})}\cos{(v_Fp\tilde{r}})}{\tilde{r}^n}.
		\end{eqnarray}
		The leading divergence is proportional to $\Lambda^2$, which is the same as that obtained in~\cite{PhysRevB.103.155151}.

	\end{widetext}

	\section{RKKY interaction in a two band lattice  model of Weyl semimetal}\label{Append:lattice model}
	To further confirm our analytical predictions, we calculate the Weyl-mediated RKKY interaction using a lattice model obtained by regularizing the continuum Hamiltonian with two Weyl nodes, %
	\begin{eqnarray}
		H=\sum_{\mathbf{k}}d_{\mu,\mathbf{k}}\sigma^\mu,
	\end{eqnarray}
	where where $\mu=0,1,2,3$ and $\sigma^0$ is the identity matrix, $\sigma^{i}$ with $i=1,2,3$ are Pauli matrices, $d_0=-t_0\sin{k_z}$, $d_1=-t\sin{k_x}$, $d_2=-t\sin{k_y}$, $d_3=M-t_z\cos{k_z}-t_p(\cos{k_x}+\cos{k_y})$, and
	the lattice constant is taken to be unity. 
	This  model has been realized in a ultracold quantum gas~\cite{Weyl_coldatom}. 
	The energy is $\epsilon_{\pm,\mathbf{k}}=d_0\pm d_{\mathbf{k}}$ with $d_{\mathbf{k}}=\sqrt{d^2_{1}+d^2_2+d^2_3}$.  The Weyl nodes are located at $\pm \mathbf{Q}=(0,0,\pm \arccos{(M-2t_p)/t_z})$. A nonzero $t_0$ induces an energy difference between the two Weyl nodes. According to our convention, the chirality of the $\pm\mathbf{Q}$ node is $\mp 1$. 
	
	The Matsubara Green's function is
	\begin{eqnarray}
		G(i\omega_n,\mathbf{k})
		=\frac{i\omega_n-d_0+\mu+\mathbf{d}\cdot\bm{\sigma}}{(i\omega_n-d_0+\mu)^2-d^2}.
	\end{eqnarray}
	In the real space,
	\begin{eqnarray}
		G(i\omega_n,\mathbf{r})=\sum_{\mathbf{k}}	G(i\omega_n,\mathbf{k})e^{i\mathbf{k}\cdot\mathbf{r}}\equiv G_\mu(i\omega_n,\mathbf{r})\sigma^\mu,
	\end{eqnarray}
	which satisfies $G^\dag(i\omega_n,\mathbf{r})=G(-i\omega_n,-\mathbf{r})$, so the RKKY interaction can be simplified as 
	\begin{eqnarray}
		\mathcal{J}_{\alpha\beta}
		=2TJ^2\mathrm{Re}\sum_{\omega_n>0}\mathrm{tr}[\sigma_\alpha G(i\omega_n,\mathbf{r})\sigma_\beta G(i\omega_n,-\mathbf{r})].
	\end{eqnarray}
	\begin{figure}[h]
		\includegraphics[width=0.45\textwidth]{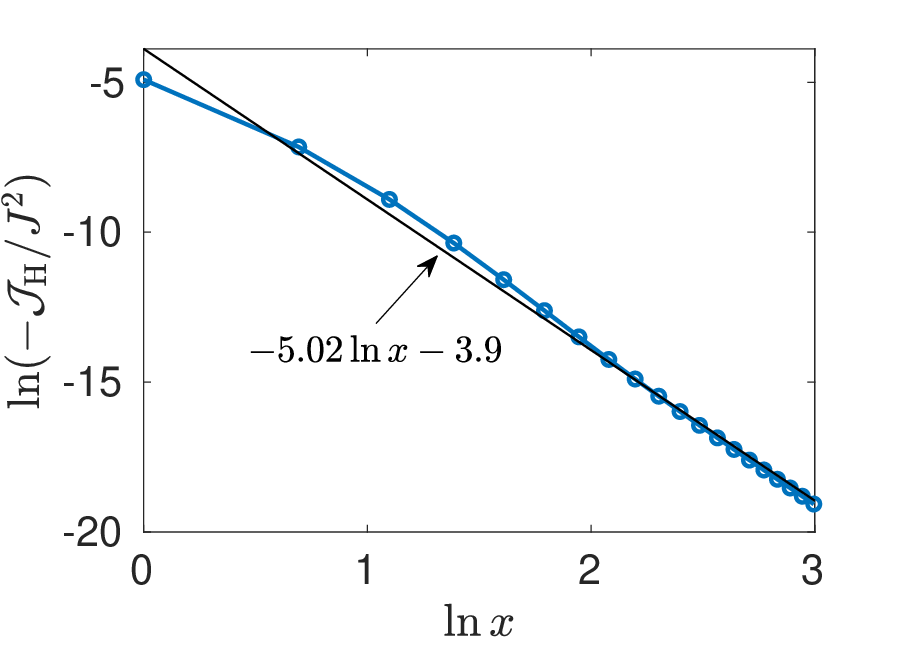}
		\caption{Double log plot of the Heisenberg interaction strength as a function of $x$. The black line is a linear fit of the data, indicating a $1/x^5$ decay of the interaction strength. We take $t=t_z=t_p=1$ as energy unit, other parameters are $M=2.5t$ and $t_0=\mu=0$. The temperature is taken to be $0.01t$, and the maximal $n$ in Matsubara frequency summation is $2000$.
		}\label{Fig:lattice_Heisenberg}
	\end{figure}
	\begin{figure}
		\includegraphics[width=0.45\textwidth]{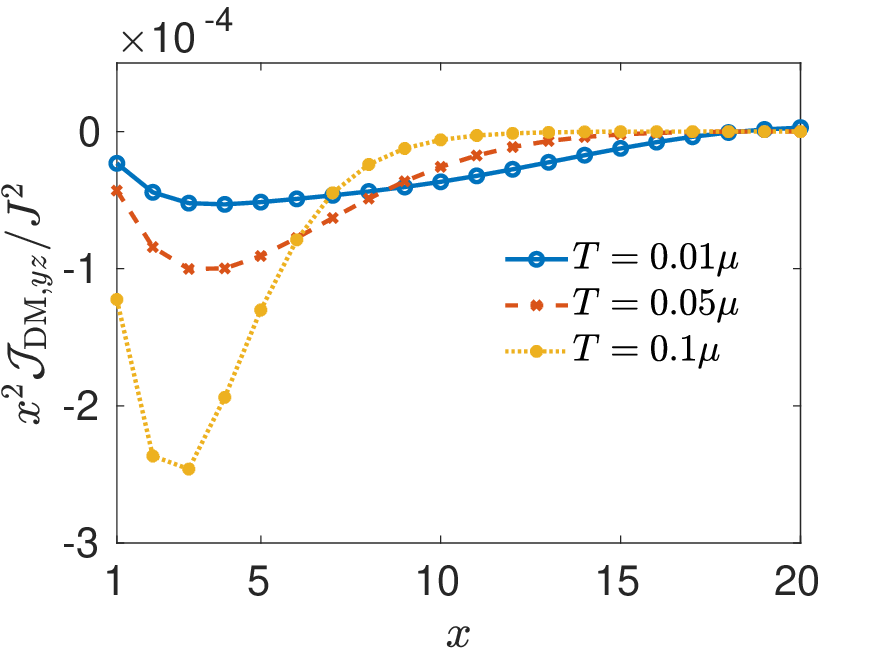}
		\caption{The DM interaction scaled by $x^2$ as a function of $x$ at several temperatures. We take $t_0=0.1t$, $\mu=-\sqrt{3}/2t_0$, and the other hopping parameters are the same as those in Fig.~\ref{Fig:lattice_Heisenberg}. The maximal $n$ in Matsubara frequency summation is $2000$.
		}\label{Fig:lattice_DMx}
	\end{figure}
	
	To examine the period and long distance behavior of the interaction, we first calculate the Heisenberg range function in the absence of the $t_0$ term and at charge neutrality. To avoid the modulation induced by Weyl nodes separation, we assume that the two magnetic moments are separated in the $x$ direction. We find
	\begin{widetext}
		\begin{eqnarray}
			\mathcal{J}_{\mathrm{H}}(x)=-4TJ^2\sum_{\omega_n>0}\bigg[\bigg(\sum_{\mathbf{k}}\frac{\omega_n\cos{(k_x x)}}{\omega^2_n+d^2_\mathbf{k}}\bigg)^2+
			\bigg(\sum_{\mathbf{k}}\frac{d_1\sin{(k_x x)}}{\omega^2_n+d^2_\mathbf{k}}\bigg)^2
			+\bigg(\sum_{\mathbf{k}}\frac{d_3\cos{(k_x x)}}{\omega^2_n+d^2_\mathbf{k}}\bigg)^2\bigg].\label{Eq:appendix_Heisenberg}
		\end{eqnarray}
	\end{widetext}

	It is evident that the Heisenberg interaction is ferromagnetic. Figure~\ref{Fig:lattice_Heisenberg} shows $\ln(-\mathcal{J}_{\mathrm{H}}/J^2)$ as a function of $\ln(x)$. As can be seen, the interaction strength is indeed non-oscillating and decays as $1/r^5$, confirming our analytical results.

	To confirm the thermal enhancement effect, we introduce a nonzero $t_0$ to generate a chiral imbalance and calculate the DM interaction at nonzero chemical potential. After some algebra, we find
	\begin{eqnarray}\label{Eq:JDM_x}
		\mathcal{J}_{\mathrm{DM},\alpha\beta}(x)
		=8\epsilon_{\alpha\beta x}TJ^2\mathrm{Im}\sum_{\omega_n>0}G_0(i\omega_n,x)G_x(i\omega_n,x).~~
	\end{eqnarray}

	Figure~\ref{Fig:lattice_DMx} shows the numerical results at different temperatures.
	The hopping parameters are the same as those in Fig.~\ref{Fig:lattice_Heisenberg}, 
	except that we take $t_0=0.1t$ and $\mu_0=-t_0\sqrt{3}/2$. The parameters are chosen such that the effective chemical potential for the Weyl node with $\chi=-1$ vanishes. Therefore, in the low energy description, only the $\chi=1$ node contributes to the DM interaction. The thermal enhancement can be clearly observed for $x\lesssim 5$. The effect is more prominent than the prediction of the continuum model.

	\bibliography{RKKY.bib}
	
\end{document}